\documentclass[natbib]{svmult}

\usepackage{mathptmx}       % selects Times Roman as basic font
\usepackage{helvet}         % selects Helvetica as sans-serif font
\usepackage{courier}        % selects Courier as typewriter font
\usepackage{type1cm}        % activate if the above 3 fonts are
\usepackage{makeidx}         % allows index generation
\usepackage{graphicx}        % standard LaTeX graphics tool
\usepackage{multicol}        % used for the two-column index
\usepackage[bottom]{footmisc}% places footnotes at page bottom

\makeindex             % used for the subject index
\begin{document}

\title*{The astrometric/morphological variability and the birth place of LS~5039}
\author{J. Mold\'on, M. Rib\'o and J. M. Paredes}
\institute{J. Mold\'on \and M. Rib\'o \and J. M. Paredes \at Departament d'Astronomia i Meteorologia and Institut de Ci\`encies del Cosmos (ICC), Universitat de Barcelona (UB/IEEC), Mart\'i i Franqu\`es 1, 08028, Barcelona, Spain,\\ 
\email{jmoldon@am.ub.es, mribo@am.ub.es, jmparedes@ub.edu}}

\maketitle

\abstract*{LS~5039 is one of the few X-ray binaries detected at VHE, and potentially contains a young non-accreting pulsar. The outflow of accelerated particles emitting synchrotron emission can be directly mapped with high resolution radio observations. The morphology of the radio emission strongly depends on the properties of the compact object and on the orbital parameters of the binary system. We present VLBA observations of LS~5039 covering an orbital cycle, which show morphological and astrometric variability at mas scales. On the other hand, we discuss the possible association of LS~5039 with the supernova remnant SNR~G016.8$-$01.1.}

\abstract{LS~5039 is one of the few X-ray binaries detected at VHE, and potentially contains a young non-accreting pulsar. The outflow of accelerated particles emitting synchrotron emission can be directly mapped with high resolution radio observations. The morphology of the radio emission strongly depends on the properties of the compact object and on the orbital parameters of the binary system. We present VLBA observations of LS~5039 covering an orbital cycle, which show morphological and astrometric variability at mas scales. On the other hand, we discuss the possible association of LS~5039 with the supernova remnant SNR~G016.8$-$01.1.}

\section{The binary system LS~5039}\label{binary}

LS~5039 is a high mass X-ray binary containing a compact object of unknown mass (1.5--10~$M_\odot$) that orbits an O6.5\,V((f)) star every 3.9~days. The system has an eccentric orbit ($e=0.35$), and it is located at 2.5~kpc \citep{casares05,aragona09}. LS~5039 is one of the few known binaries that displays high-energy (HE) and very high energy (VHE) gamma-ray emission ($E>100$~GeV) with a clear orbital modulation \citep{aharonian06, abdo09}. Several theoretical models have been developed to explain the multiwavelength orbital behavior of LS~5039. The HE/VHE emission is basically interpreted as the result of inverse Compton upscattering of stellar UV photons by relativistic electrons. The acceleration of electrons seems to be produced by shocks between the relativistic wind of a young non-accreting pulsar and the wind of the stellar companion \citep{dubus06, sierpowska08}. No radio pulses have been detected at 1.4~GHz \cite{morris02}, although the strong free-free absorption with the stellar wind may prevent any detection of pulsed radio emission. Therefore, the nature of the compact object (black hole or pulsar) is unknown.

In the non-accreting pulsar scenario the shocked material is contained by the stellar wind behind the pulsar, producing a 'bow' shaped nebula extending away from the stellar companion. The high energy emission is produced in the region where the wind pressures are balanced, while a tail of accelerated particles forms behind the pulsar. The cooling processes of these accelerated particles along the adiabatically expanding flow produce the non-thermal broadband emission from radio to X-rays. The morphology is similar to the one expected in isolated pulsars moving through the ISM but, as a consequence of the orbital motion of the binary system, the tail of the flow bends following an elliptical path during the orbital cycle.

\section{Study of the milliarcsecond radio emission of LS~5039} \label{testing}

Free-free absorption on the stellar wind becomes optically thin at distances $\sim1$~AU. At scales of a few AU the radio structure produced by the outflow becomes relevant. LS~5039 is located at 2.5~kpc, and hence the morphology can be studied at scales of 1--10~mas (2.5--25~AU), which are directly observable with VLBI. In this context, the scenario provides two main predictions about the radio behavior at mas scales. On one hand, it is expected that the direction of the extended emission changes with the pulsar's orbital motion. On the other hand, the peak of the radio emission should trace an ellipse of a few mas \citep{dubus06}. The orbital changes of the emission at different frequencies contain relevant information on the nature of the compact object, the orbital parameters of the system, the geometry of the region where particles are accelerated, and the processes involved. 

\section{Radio observations} \label{radio}

The first VLBA observation of LS~5039 at 5 GHz, conducted in 1999, showed a $\sim$20~mJy central core and bipolar extended emission of a few mJy extending over 6~mas on the plane of the sky \citep{paredes00}. Similar observations at two different orbital phases \citep{ribo08} showed a changing morphology at mas scales: the core component had a constant flux density, and the Position Angle (P.A.) of the direction of the elongated emission changed by $12\pm3^{\circ}$ between both runs, with a remarkable symmetry change. Therefore, morphological changes at 5~GHz occur on timescales of the order of the orbital period.

To further study the morphological variability, we observed the source in 2007 during five consecutive days, to cover a full orbit of 3.9~days, and an extra day to disentangle between orbital or secular variability. The observations were centered at the orbital phases 0.98, 0.23, 0.49, 0.75, and 0.00, each one spanning 6~hours (0.07 in phase), computed using the ephemerides of \cite{aragona09}. The observations were phase-referenced to the phase calibrator J1825$-$1718, and an astrometric check source was observed regularly during the runs. The self-calibrated images show a main core and extended emission up to $\sim6$~mas, as in previous observations. The P.A.\ of the extended emission from the main core changes significantly every day, covering angles between $-$50 and $-85^{\circ}$ for phases in the range 0.5--0.0, and P.A.\ 120$^\circ$ at phase 0.23. Therefore, a subtle change in morphology happens in less than 20~hours after periastron. 

The resolution of the phase-referenced images is limited by the scatter broadening of the calibrator. The astrometric accuracy, extrapolated from the mean dispersion of the astrometric check source, is 0.23 and 0.27~mas in right ascension and declination, respectively (Mold\'on et al., in preparation). The positions of the peak of the emission for phases 0.5--0.0 are compatible within errors. The peak of the emission at phase 0.23 shows a displacement of $2.3\pm0.3$~mas in P.A.\ $26^{\circ}$, nearly opposite to the extended emission sense. This indicates that the variations of the extended emission with respect to the core are a combination of intrinsic variability of this extended emission and an absolute displacement of the core component.

\section{Proper motion} \label{Proper motion}

\begin{figure}[ht]
\begin{center}
\sidecaption
\includegraphics[scale=.38]{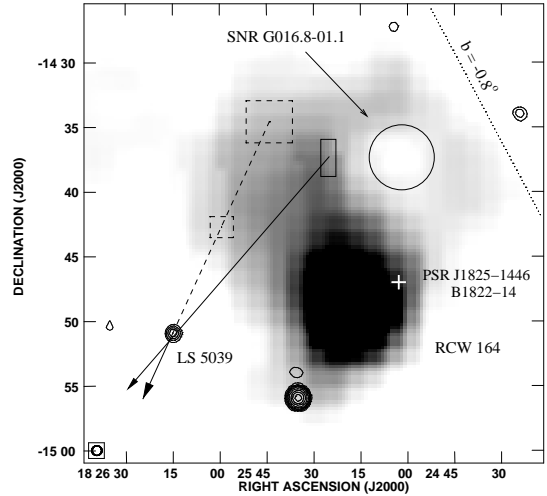}
\caption{Wide field radio map of LS~5039, showing the position of the nearby SNR~G016.8$-$01.1. The grey scale emission is at the 6~cm wavelength. The overlaid contours correspond to the NVSS map at the 20~cm. The arrows mark the proper motion senses. The dashed line is the computed trajectory reported in Rib\'o et al. (2002), whereas the solid line is the computed trajectory from our fit (see text). The error boxes are calculated for 10$^5$~yr ago.}
\label{fig:1}       % Give a unique label
\end{center}
\end{figure}

As a by-product, these phase-referenced observations provide one average precise position of the source in the sky. \cite{ribo02} computed the trajectory of LS 5039 for the last $10^5$~yr using optical/radio astrometry from 1905 to 2002. Their result marginally suggests an association with SNR~G016.8$-$01.1. This new VLBA position, combined with previous radio interferometric measurements spanning 9 years, allow us to calculate a more precise proper motion and past trajectory of the source (see solid line in Fig.~\ref{fig:1}). The new past trajectory of LS~5039 is compatible with the center of the SNR. With a firm association, a kinematical age of the compact object would be obtained, which would have direct implications on its properties. In case the system contains a pulsar, it would probably be a young pulsar in the non-accreting phase, and therefore, compatible with the shocked winds scenario. 

%------------------------------------------------------------------------------

%------------------------------------------------------------------------------

\section{Discussion} \label{discussion}

The emission at 5~GHz traces the short-lived electrons of the outflow up to 15~AU. The morphological changes along the orbit allow us to model the velocity, the energy, and the cooling time-scale of the flow of particles that originates the extended emission. The morphological and astrometric information constrains the inclination of the orbit ($i$), a key parameter of the system. The mass function of the system and $i$ yield the mass of the compact object. On the other hand, it should be possible to clearly trace the peak position along the orbit with observations at higher frequencies (where the phase calibrator is much more compact), yielding direct information on the absorption around the system. 

The VLBA images from 2007 at 5~GHz contain relevant information to test the models (Mold\'on et al.\ in prep.), but the lack of continous astrometric information following the peak of the emission at all orbital phases is a key point to better constrain the physical properties of the system. If the peak position is expected to be shifted between 1--2~mas, it is not possible to unambiguously distinguish the displacement of the peak from intrinsic variations of the extended emission. The flow velocity and cooling times strongly depend on this ambiguity, which can be disentangled with accurate astrometry, to be obtained in the future.

\bibliographystyle{}
%\bibliography{biblio}
%%%%%%%%%%%%%%%%%%%%%%%% referenc.tex %%%%%%%%%%%%%%%%%%%%%%%%%%%%%%
% sample references
% %
% Use this file as a template for your own input.
%
%%%%%%%%%%%%%%%%%%%%%%%% Springer-Verlag %%%%%%%%%%%%%%%%%%%%%%%%%%
%
% BibTeX users please use
% \bibliographystyle{}
% \bibliography{}
%
\def\aap{A\&A}%
          % Astronomy and Astrophysics
\def\aj{AJ}%
          % Astronomical Journal
\def\apj{ApJ}%
          % Astrophysical Journal
\def\apjl{ApJ}%
          % Astrophysical Journal, Letters
\def\mnras{MNRAS}%
          % Monthly Notices of the RAS

\end{document}